\RequirePackage[hyphens]{url}
\documentclass[twocolumn,showpacs,preprintnumbers,superscriptaddress]{revtex4-1}

\usepackage{color, amsmath, amssymb, graphicx, float, dcolumn, bm}

\usepackage{revsymb,natbib,placeins,setspace}
\usepackage{hyperref}
\PassOptionsToPackage{hyphens}{url}\usepackage{hyperref}

\begin{document}

\title{Steady-State Control and Machine Learning of Large-Scale Deformable Mirror Models}

\author{Aleksandar Haber} 
\affiliation{Department of Engineering and Environmental Science, The City University of New York, College of Staten Island, New York, NY 10314, USA, aleksandar.haber@csi.cuny.edu} 
 \begin{abstract}
We use Machine Learning (ML) and system identification validation approaches to estimate neural network models of large-scale Deformable Mirrors (DMs) used in Adaptive Optics (AO) systems. To obtain the training, validation, and test data sets, we simulate a realistic large-scale Finite Element (FE) model of a faceplate DM. The estimated models reproduce the input-output behavior of Vector AutoRegressive with eXogenous (VARX) input models and can be used for the design of high-performance AO systems. We address the model order selection and overfitting problems. We also provide an FE based approach for computing steady-state control signals that produce the desired wavefront shape. This approach can be used to predict the steady-state DM correction performance for different actuator spacings and configurations. The presented methods are tested on models with thousands of state variables and hundreds of actuators. The numerical simulations are performed on low-cost high-performance graphic processing units and implemented using the TensorFlow machine learning framework. The used codes are available online. The approaches presented in this paper are useful for the design and optimization of high-performance DMs and AO systems. 
\end{abstract} 
\maketitle  
To develop high-performance Adaptive Optics (AO) systems it is of paramount importance to accurately model the main components, such as Deformable Mirrors (DMs), WaveFront Sensors (WFSs), and other physical processes that affect the system performance~\cite{tyson2010principles, Haber:13, sinquin2018tensor,vogel2010modeling,Fernandez2003}. The main focus of this paper is on estimation and steady-state control of large-scale DMs. The large-dimensionality of DMs manifests itself in two forms. First of all, due to the continuous effort to increase the performance of AO systems, numbers of DM actuators constantly increase. Thus, in AO systems for extremely large telescopes, DMs have hundreds or even thousands of spatially distributed actuators. Secondly, even if a number of DM actuators is modest, a DM is inherently a large-scale system, since its dynamics is mainly governed by infinite-dimensional Partial Differential Equations (PDEs). For example, to describe the dynamics of faceplate DMs, we need to employ the Kirchhoff-Love theory of plates~\cite{timoshenko1959theory}. On the other hand, the dynamics of thermally actuated DMs are described by thermoelastic PDEs~\cite{haberThesis,haber2013identification,haber2013predictive}. To accurately describe large-scale DM dynamics, we need state-space~\cite{haber2018sparsity} or Vector AutoRegressive with eXogenous (VARX) input models of high model orders. 
When all the system parameters are known accurately, we can employ Finite Element (FE) discretization techniques~\cite{zienkiewicz2000finite} to develop DM  models. However, in practice, it if often the case that the system parameters are only partially known or completely unknown. In such cases, we need to use system identification techniques~\cite{verhaegen2007filtering,haber2014subspace} to estimate models from the experimental data. In~\cite{haber2013identification,haber2013predictive,chiuso2009dynamic,song2011controller} it has been demonstrated that Subspace Identification (SI) techniques are able to estimate reduced-order DM models. Recently, Machine Learning (ML) Neural Network (NN) methods \cite{goodfellow2016deep} became a viable alternative to SI methods for estimating large-scale models of dynamical systems~\cite{haber2019comsol,haber2019subspaceASME}. This is mainly due to the rapid developments of parallel computing technologies and devices, such as NVIDIA  Graphic Processing Units (GPUs) and the CUDA computing platform.  In this Letter, we combine ML NN techniques with system identification model validation approaches to estimate NN models of large-scale thin faceplate DMs. Since we do not have access to the experimental data, a realistic FE model of a DM is used as a data-generating model. By simulating this model for random initial conditions and control inputs, we generate input-output data that is used to train the NN. We address the model selection and overfitting problems. Our results show that shallow NN can effectively estimate DM models with hundreds of actuators. The approach used in this paper can also be used to build off-line pre-trained models whose parameters can be quickly adapted on the basis of the new experimental data. Pre-trained generic models are known to significantly speed up the learning process~\cite{goodfellow2016deep}. 
Besides this,  we also provide an FE based approach for computing steady-state control signals that produce the desired wavefront shape. This approach can be used to predict the steady-state DM correction performance for different actuator spacings and configurations. The presented methods are implemented in TensorFlow and Keras~\cite{cholletBook} on a low-cost high-performance GPU (GeForce RTX 2080 Ti) and the codes are available online~\cite{haberAOcode2019}. The results of this paper and the developed codes  provide useful numerical tools for improving the performance of large-scale DMs and AO systems.

\begin{figure}[H]
\centering 
\includegraphics[scale=0.43,trim=0mm 0mm 0mm 0mm ,clip=true]{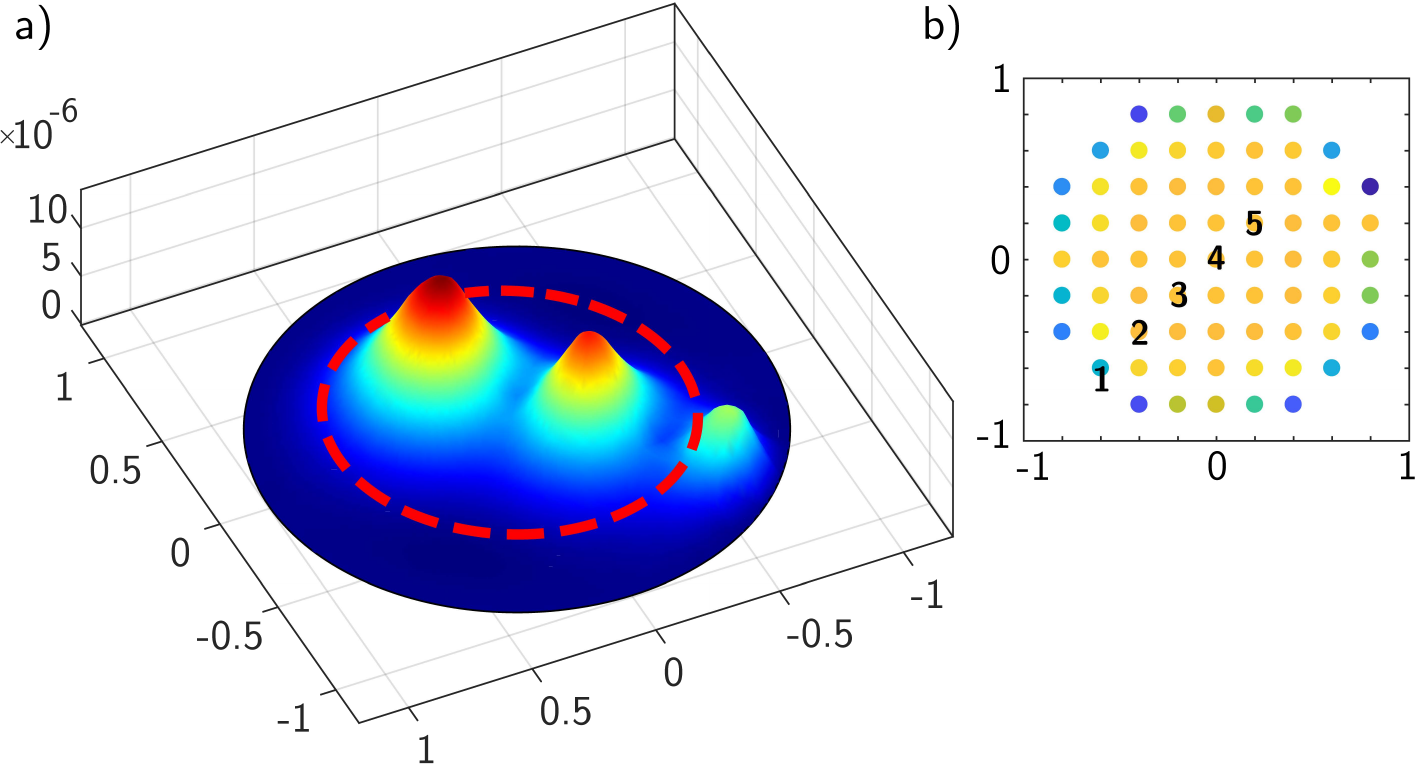}
\vspace{-3mm}
\caption{(a) The global deformation of the mirror surface when the actuators $1-5$ in the panel (b) are active. The red dashed line encircles the observed deformation area of the DM.}
\label{fig:Graph1}
\end{figure}

We start by describing a model of a face plate DM originally developed in~\cite{haber2019DMpaper}. The mirror consists of a thin face plate that is deformed by a rectangular array of actuators. The face plate has a radius of $1 [m]$ and thickness of $0.003 [m]$. It is made of Zerodur\textsuperscript{\textregistered} with the Young's modulus, density, Poisson's ratio of $9.03  \cdot 10^{10} [Pa]$,  $2530$  $[kg/m^{3}]$, and $0.24$, respectively. The actuators are modeled as mass-spring-damper systems with the stiffness, damping, and mass  of $10^{4}$ [N/m], $500$ [Ns/m], and $0.3 [kg]$, respectively. The state-space model is obtained by using the FE method and the LiveLink\textsuperscript{\textregistered} for MATLAB COMSOL Multiphysics\textsuperscript{\textregistered} package. Figure~\ref{fig:Graph1}(a) shows the simulated steady-state DM deformation when the actuators $1,2,3,4,$ and $5$ in Fig.~\ref{fig:Graph1}(b) are active (actuator spacing of $0.2\;[m]$). The forces of the actuators $1,3,$ and $5$ are $1.5$, $1$, and $1$ $[N]$, respectively, while the forces of the actuators $2$ and $4$ are $-0.5$ $[N]$.
\vspace{-2mm}
\vspace{-5mm}
We assume that the deformation of the area of the radius $0.6 [m]$ is observed by a WFS or an interferometer. This area is encircled by a dashed line in Fig.~\ref{fig:Graph1}(a). We assume that only out-of-plane deformations can be observed. The resulting FE model together with the measurement (observation) equation has the following form 
\begin{align}
M_{1}\ddot{\mathbf{z}}+M_{2}\dot{\mathbf{z}}+M_{3}\mathbf{z}=B\mathbf{u}, \;\; \mathbf{y}=C\mathbf{z},
\label{FEmodel}
\end{align}
where $M_{1}\in \mathbb{R}^{n\times n}$, $M_{2}\in \mathbb{R}^{n\times n}$,  and $M_{3}\in \mathbb{R}^{n\times n}$ are the mass, damping, and, stiffness matrices, $B\in \mathbb{R}^{n\times m}$, is the control input (forces) matrix, $C\in \mathbb{R}^{r\times n} $ is the measurement matrix, $\mathbf{z}=\mathbf{z}(t)\in \mathbb{R}^{n}$ is the displacement vector, $\mathbf{u}\in \mathbb{R}^{m}$ is the control input vector (consisting of actuator forces), and $\mathbf{y}\in \mathbb{R}^{r}$ is the output vector (observed deformations). 

\begin{figure}[H]
\centering 
\includegraphics[scale=0.33,trim=0mm 0mm 0mm 0mm ,clip=true]{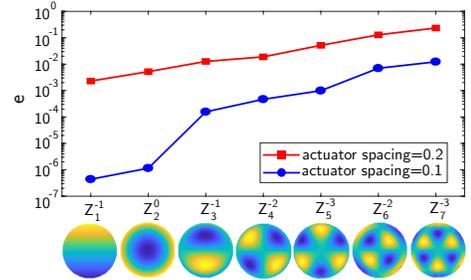}
\vspace{-3mm}
\caption{The steady-state wavefront error. The notation $Z_{i}^{j}$ denotes the Zernike polynomials.}
\label{fig:Graph2}
\end{figure}

\begin{figure}[H]
\centering 
\includegraphics[scale=0.40,trim=0mm 0mm 0mm 0mm ,clip=true]{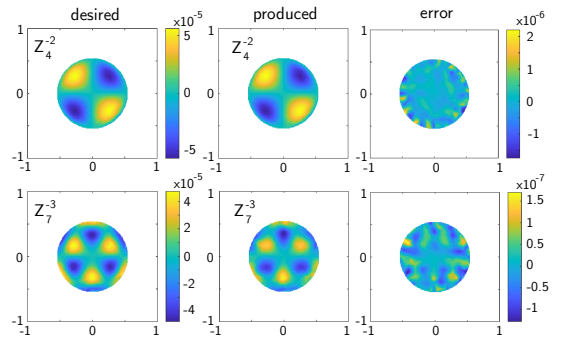}
\vspace{-3mm}
\caption{The desired and produced wavefronts and the global error. The results are generated for the actuator spacing of $0.2$.}
\label{fig:Graph3}
\end{figure}

\begin{figure}[H]
\centering 
\includegraphics[scale=0.27,trim=0mm 0mm 0mm 0mm ,clip=true]{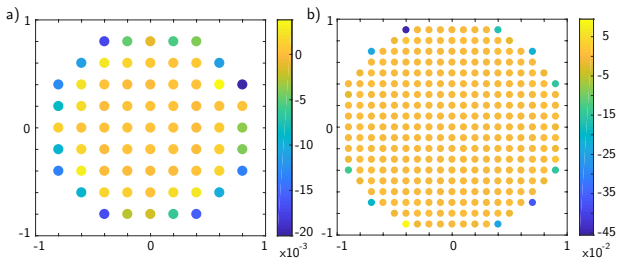}
\caption{The control forces for the actuator spacing of (a) $0.2$ and (b) $0.1$. The control forces are obtained for the desired wavefront shape $Z_{2}^{0}$.}
\label{fig:Graph4}
\end{figure}

Our first goal is to determine the control input vector that produces a desired steady-state DM deformation (wavefront shape). In the steady-state, we have $\ddot{\mathbf{z}}=0$ and $\dot{\mathbf{z}}=0$, and from the mirror model in \eqref{FEmodel}, we obtain $M_{3}\mathbf{z}=B\mathbf{u}$. Then, given a desired mirror surface profile, expressed by $\mathbf{y}_{d}$, the problem is to determine $\mathbf{u}$ that will produce such a profile in steady-state. A naive approach for solving this problem would be to express the state vector as follows: $\mathbf{z}=M_{3}^{-1}B\mathbf{u}$, and then to substitute this expression in $\mathbf{y}_{d}=C\mathbf{z}$. The resulting expression has the following form $\mathbf{y}_{d}=CM_{3}^{-1}B\mathbf{u}$. This equation can be solved for $\mathbf{u}$ in the least-squares sense~\cite{verhaegen2007filtering}. The matrix $CM_{3}^{-1}B$ can be seen as the DM influence function~\cite{tyson2010principles}. There are two problems with such an approach. First of all, although the matrix $M_{3}$ is sparse, its inverse is completely dense, and consequently, for large-scale problems  it cannot be computed and stored in a computer memory. Furthermore, the matrix $M_{3}$ can be ill-conditioned, and consequently, its inverse is prone to numerical errors. Motivated by this, we simultaneously solve $M_{3}\mathbf{z}=B\mathbf{u}$ and $\mathbf{y}_{d}=C\mathbf{z}$ by solving the following system for the vector $\mathbf{w}$
\begin{align}
& \mathbf{g}=S\mathbf{w},  \;\; S=\begin{bmatrix} M_{3} & -B \\ C & 0 \end{bmatrix},   \mathbf{g}=\begin{bmatrix} 0  \\  \mathbf{y}_{d}  \end{bmatrix}, \;\; \mathbf{w}=\begin{bmatrix} \mathbf{z} \\  \mathbf{u} \end{bmatrix}.    \label{augmentedSystem}  
\end{align}
For certain selections of the vector $\mathbf{y}_{d}$, the system in \eqref{augmentedSystem} might not have a solution. Consequently, we solve the system in the least-squares sense:
\begin{align}
\min_{\mathbf{w}} \left\|  \mathbf{g}-S\mathbf{w}  \right\|_{2}^{2}.
\label{LSformulation}
\end{align} 
We solve the problem in \eqref{LSformulation} using the multifrontal sparse QR factorization  method~\cite{davis2016survey}. Besides being used for control, the presented approach can be useful for predicting the DM performance for various actuator spacings or for optimizing the system performance. Figure~\ref{fig:Graph2} shows the error of producing desired wavefront shapes (Zernike polynomials denoted by $Z_{i}^{j}$ ~\cite{tyson2010principles}) for the actuator spacing of $0.1$ and $0.2$. For the actuator spacing of $0.1$, we have $r=1724$ (out-of-place displacements at the observed points), $n= 14055$ (model dimension), and $m=305$ (actuator number). For the actuator spacing of $0.2$, we have $r=308$, $n=2439$, and $m=69$.  The control error, quantifying the correction performance,  is defined by  $e=\left\|\mathbf{y}_{d}- \mathbf{y}^{*}  \right\|_{2}/\left\| \mathbf{y}_{d} \right\|_{2}$, where  $\mathbf{y}^{*}$ is the wavefront produced by the computed control input. From Fig.~\ref{fig:Graph2}, we observe that by doubling the actuator spacing we achieve at least an order of magnitude error improvement. 
\vspace{-2mm}
\vspace{-2mm}
Figure~\ref{fig:Graph3} shows the global wavefront shapes and the global errors for the actuator spacing of $0.2$. Figure~\ref{fig:Graph4} shows the control forces. 
\vspace{-2mm}
Next, we explain methods for estimating and validating DM models using the NN ML techniques. To reduce the model output dimensionality, we express the system output using the Zernike basis functions~\cite{tyson2010principles}. Let $\mathbf{q} \in \mathbb{R}^{l}$ be a vector of the first $l$ Zernike coefficients (excluding the piston term), where $l\ll r$. This is a new system output that can be expressed as $\mathbf{q}=C_{1}\mathbf{y}$, where $C_{1}\in \mathbb{R}^{l \times r}$ is the matrix mapping deformations into the Zernike coefficients. Because in practice models are estimated using discrete-time data, we formulate the estimation problem in the discrete-time domain. We consider system's inputs and new outputs at the discrete-time instants $t=k h$, where $h$ is a small \textit{discretization step}, and $k=0,1,2,\ldots$ is a \textit{discrete-time instant}. Let $\mathbf{q}_{k}$ and $\mathbf{u}_{k}$ denote the system output and input, respectively, at the discrete-time instant $k$. Then, by writing the model in \eqref{FEmodel} in the state-space form, and by discretizing such a model using the backward Euler method~\cite{haber2019DMpaper}, and by using elementary control-theory arguments~\cite{haber2019subspaceASME}, the system output can be expressed as follows
\begin{align}
\mathbf{q}_{k} = \sum_{i=1}^{p}Q_{i} \mathbf{q}_{k-i} + \sum_{i=1}^{p}U_{i} \mathbf{u}_{k-i}+\mathbf{e}_{k},
\label{VARXmodel}
\end{align}
where $Q_{i}\in \mathbb{R}^{l\times l}$, $U_{i}\in \mathbb{R}^{l \times m}$, $\mathbf{e}_{k} \in \mathbb{R}^{l}$ is the measurement noise vector, and $p$ is the \textit{past window}. The model in \eqref{VARXmodel} is the VARX model which states that the system output at the time instant $k$ is a linear combination of past outputs, inputs, and the current measurement noise. This VARX model motivates us to postulate an NN model having the following form
\begin{align}
\hat{\mathbf{q}}_{k}=N(\mathbf{q}_{k-1},\ldots, \mathbf{q}_{k-p},\mathbf{u}_{k-1},\ldots, \mathbf{u}_{k-p}),
\label{neuralNetwork}
\end{align}
where $\hat{\mathbf{q}}_{k}$ is the NN output which an estimate of the system output, and the NN inputs are the vectors $\mathbf{q}_{k-1},\ldots, \mathbf{q}_{k-p}$ and $\mathbf{u}_{k-1},\ldots, \mathbf{u}_{k-p}$. The notation $N(\cdot)$ denotes an NN function mapping NN inputs into outputs. For notation brevity, we denote the NN by a single function $N(\cdot)$ which is a composition of activation functions. Namely, the NN consists of hidden layers that map inputs into outputs and every layer consists of a number of neurons, for more details, see~\cite{cholletBook,goodfellow2016deep}. The vector $\hat{\mathbf{q}}_{k}$ is referred to as the \textit{closed-loop prediction}. On the other hand, the vector $\tilde{\mathbf{q}}_{k}$ defined by
\begin{align}
\tilde{\mathbf{q}}_{k}=N(\tilde{\mathbf{q}}_{k-1},\ldots, \tilde{\mathbf{q}}_{k-p},\mathbf{u}_{k-1},\ldots, \mathbf{u}_{k-p}),
\label{neuralNetworkOpenLoop}
\end{align}
is also an estimate of the system output and is referred to as the \textit{open-loop} prediction. In contrast to the closed-loop prediction, the open-loop prediction is computed on the basis of the past estimates of the system output. The estimation problem is to estimate the past window $p$, and the parameters (weights) of the NN defined by \eqref{neuralNetwork}, such that the predicted output $\hat{\mathbf{q}}_{k}$ accurately predicts the true system output $\mathbf{q}_{k}$. In mathematical terms, the estimation problem is to minimize the Mean Squared Error (MSE) between the prediction sequence $\hat{\mathbf{q}}_{i}$ and the true output sequence $\mathbf{q}_{i}$ for $i=p,p+1,\ldots, f$.

We use the TensorFlow and Keras ML Python libraries~\cite{cholletBook} to train the NN. On the basis of the step-response analysis, we choose $h=1\cdot 10^{-3}\; [s]$ (for more details see~\cite{haber2019DMpaper}). The network has $2$ hidden layers. Each layer has $32$ neurons. We use linear activation functions and train the network for $45000$ epochs. Data sets consist of  $f=4000$ samples of inputs and outputs. We generate three data sets for three different initial displacements and control input sequences. In each case, initial displacements and control input sequences are generated as Gaussian white noise. The three data sets are referred to as the \textit{training, validation}, and \textit{test data sets}. The training data set is used to fit the NN parameters. While fitting the parameters, in every iteration (epoch), using the validation data set, we compute the MSE validation value on the basis of the system outputs and NN closed-loop predictions.  For a fixed value of $p$, a final estimated model is a model that produces the smallest MSE validation value. In this way, we avoid network overfitting. Once the model has been fitted, the test data set is used to compute the \textit{final prediction performance}. We use two prediction performance measures. The \textit{closed-loop prediction performance} is expressed by: $\varepsilon_{CL}= || \underline{\mathbf{q}}-\underline{\hat{\mathbf{q}}}  ||_{2}  / || \underline{\mathbf{q}} ||_{2}$, where $\underline{\mathbf{q}}= \begin{bmatrix} \mathbf{q}_{p}^{T} \; ... \; \mathbf{q}_{f}^{T} \end{bmatrix}^{T}$ and $\hat{\underline{\mathbf{q}}}= \begin{bmatrix}\hat{ \mathbf{q}}_{p}^{T} \; ... \;\hat{\mathbf{q}}_{f}^{T} \end{bmatrix}^{T}$, where $\hat{\mathbf{q}}_{k}$, $k=p,...,f$. is computed using \eqref{neuralNetwork}. The \textit{open-loop prediction performance} is expressed by $\varepsilon_{OL}= || \underline{\mathbf{q}}-\underline{\tilde{\mathbf{q}}}  ||_{2}  / || \underline{\mathbf{q}} ||_{2}$, where $\underline{\tilde{\mathbf{q}}}= \begin{bmatrix} \tilde{\mathbf{q}}_{p}^{T} \;  ... \; \tilde{\mathbf{q}}_{f}^{T} \end{bmatrix}^{T}$, where $\tilde{\mathbf{q}}_{k}$ is an \textit{open-loop prediction} computed on the basis of \eqref{neuralNetworkOpenLoop}. We estimate the past window $p$ and consequently, the final model as follows. First, we train the network for increasing values of $p$. For every value of $p$ we obtain a single model. We compute the final prediction performances for such a network. The final value of $p$ is the one that produces the smallest open-loop prediction performance. In some cases this selection method might produce high order models. As an alternative, we also select the parameter $p$ using the Akaike Information Criterion (AIC)~\cite{lutkepohl2005new,haber2019subspace}. The AIC provides a trade-off between the goodness of the fit and the number of model parameters. For every model, on the basis of the network prediction, we compute the value of the AIC.  The final value of $p$ is the one that minimized the AIC value.

\begin{figure}[H]
\centering 
\includegraphics[scale=0.3,trim=0mm 0mm 0mm 0mm ,clip=true]{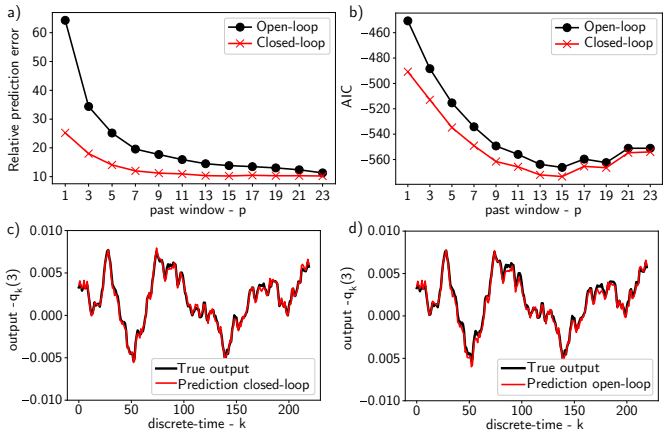}
\caption{The estimation performance for the model with the actuator spacing of $0.2$ in the noise-free case. (a) The final prediction performance. (b) The AIC value. (c) The closed-loop and (d) open loop prediction performances for $p=15$.}
\label{fig:Graph5}
\end{figure}

\begin{figure}[H]
\centering 
\includegraphics[scale=0.3,trim=0mm 0mm 0mm 0mm ,clip=true]{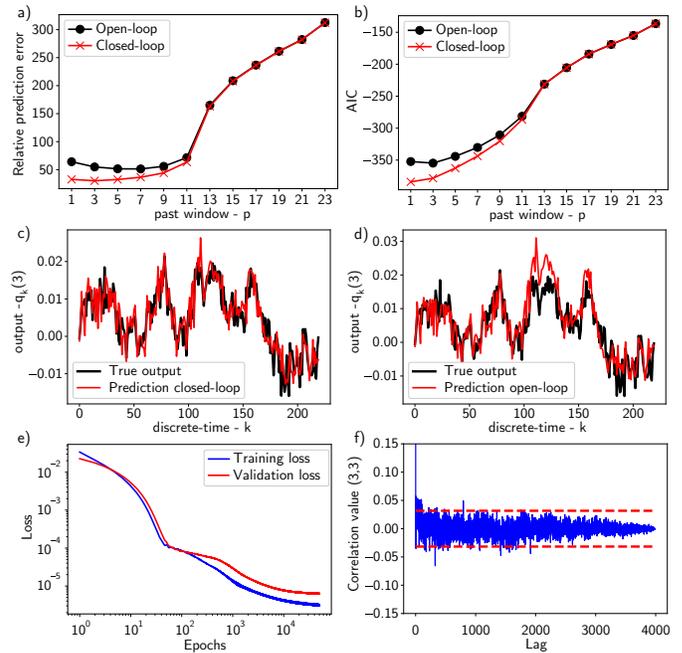}
\caption{The estimation performance for the model with the actuator spacing of $0.1$ in the case of the measurement noise. (a) The final prediction performance. (b) The AIC value. (c) The closed-loop and (d) open loop prediction performances for $p=15$. (e) The training and validation losses. (f) The residual correlation analysis. The red dashed line denotes $5 \%$ confidence interval for the white-noise hypothesis test.}
\label{fig:Graph6}
\end{figure}

First, we test the estimation performance when the system outputs are not corrupted by the measurement noise. We use a model with the actuator spacing of $0.2$ having $l=32$ outputs (total number of Zernike coefficients) and $m=69$ inputs (actuator number). The final prediction performance and AIC values are shown in Fig.~\ref{fig:Graph5}(a) and \ref{fig:Graph5}(b). Since in this case, the AIC graph is more informative, on the basis of the AIC values, we select $p=15$. Figures~\ref{fig:Graph5}(c) and (d) show the ''true'' Zernike coefficient (system output) and the predicted one in the closed-loop and open-loop prediction case. We see that in both cases, the NN is able to accurately predict the system output. Next, we add white Gaussian noise to the output vectors $\mathbf{q}_{k}$. The signal to noise ratio is $20$. We test the prediction performance for the large-scale model with the actuator spacing of $0.1$ having $l=32$ outputs and $m=305$ inputs. The results are shown in Figs.~\ref{fig:Graph6} (a)-(d) and they are analogous to the results shown in the corresponding panels of Fig~\ref{fig:Graph5}. The past window of $p=5$ is selected on the basis of the final prediction performance. To illustrate the training and validation processes, Fig.~\ref{fig:Graph6}(e) shows the training and validation MSE values (losses). This figure shows that we can additionally improve model performance by increasing the number of training epochs. As a final model quality test, we perform the residual correlation analysis~\cite{lutkepohl2005new,haber2019subspace}. The results for the output $3$ are shown in Fig.~\ref{fig:Graph6}(f). Only $1.7 \%$ of the correlation values fall outside the $5\%$ confidence interval. This indicated that the prediction error resembles a white-noise process, and since our measurement noise is white, this implies that the fitted model captures all the dominant modes of the system dynamics. 

In conclusion, in this Letter, we have presented a machine learning-based approach for estimating models of large-scale DMs having hundreds of actuators. We have presented approaches for selecting the model parameters and for validating the prediction performance. We have shown that a shallow neural network is able to accurately predict the system output. Besides this, in this paper, we have presented a steady-state control approach for DMs that is based on the FE model. This approach can be used in the DM design process to predict and optimize DM correction performance. The results of this paper and codes provide effective tools to improve the performance of large-scale DMs and AO systems.

\noindent\textbf{Funding.} This work is supported by the PSC-CUNY Award A (61303-00 49) and the PSC-CUNY Award A (62267-00 50).

\noindent\textbf{Disclosures.} The authors declare no conflicts of interest.

\bibliography{sample}
\bibliographystyle{ieeetr}

\end{document}